\documentclass[aps,prl,twocolumn,showpacs]{revtex4}
\usepackage{graphicx}
\begin{document}
\title{Determination of the parameters of semiconducting
CdF$_{2}$:In with Schottky barriers from radio-frequency
measurements}
\author{A. I. Ritus\footnote{Electronic address: ritus@ran.gpi.ru}, A. V. Pronin, and  A. A. Volkov}%
\affiliation{Institute of General Physics, Russian Academy of
Sciences, 119991 Moscow, Russia}%
\author{P. Lunkenheimer and A. Loidl}%
\affiliation{Experimentalphysik V, EKM, Universit\"{a}t Augsburg,
86135 Augsburg, Germany}%
\author{A. S. Shcheulin and A. I. Ryskin}%
\affiliation{S. I. Vavilov State Optical Institute, 195034 St.
Petersburg, Russia}
\date{\today}

\begin{abstract}
Physical properties of semiconducting CdF$_{2}$ crystals doped
with In are determined from measurements of the radio-frequency
response of a sample with Schottky barriers at frequencies $10 -
10^{6}$ Hz. The \textit{dc} conductivity, the activation energy of
the amphoteric impurity, and the total concentration of the active
In ions in CdF$_{2}$ are found through an equivalent-circuit
analysis of the frequency dependencies of the sample complex
impedance at temperatures from 20 K to 300 K. Kinetic coefficients
determining the thermally induced transitions between the deep and
the shallow states of the In impurity and the barrier height
between these states are obtained from the time-dependent
radio-frequency response after illumination of the material. The
results on the low-frequency conductivity in CdF$_{2}$:In are
compared with submillimeter ($10^{11} - 10^{12}$ Hz) measurements
and with room-temperature infrared measurements of undoped
CdF$_{2}$. The low-frequency impedance measurements of
semiconductor samples with Schottky barriers are shown to be a
good tool for investigation of the physical properties of
semiconductors.
\end{abstract}

\pacs{ 72.20.-i, 77.22.-d, 78.20.Ci }

\maketitle

\section{\label{sec:level1}Introduction}

CdF$_{2}$ belongs to a large family of fluorides crystallizing in
the cubic fluorite structure \textit{O5h} (\textit{Fm3m}), with
parameters typical for this family: it is an ionic dielectric with
a wide band gap and only one dipole-active lattice mode centered
near $6\cdot10^{12}$ Hz \cite{axe}. The latter provides an almost
frequency and temperature independent lower frequency (static)
dielectric constant $\varepsilon \cong 8$ \cite{axe,pronin}.

\begin{figure}[]
\centering
\includegraphics[width=\columnwidth,clip]{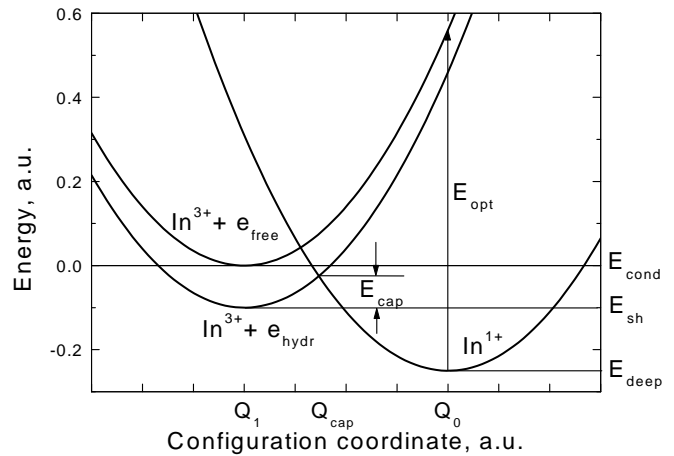}
%\vspace{0.2cm}
\caption{Schematic diagram of the energy levels for
the deep, $E_{deep}$, and the shallow, $E_{sh}$, states of the
bistable In center in CdF$_{2}$, as a function of the
configuration coordinate $Q$ of the In nucleus. $E_{cap}$ - the
barrier height between the shallow and the deep states, $E_{cond}$
- the bottom of the conduction band, $E_{opt}$ - the optical
ionization energy of the deep state. } \label{fig1}
\end{figure}

If CdF$_{2}$ crystals doped with In are heated up in a reducing
atmosphere of hydrogen or alkali metal vapors in a so-called
additive coloration process, when a part of the interstitial
F$^{1-}$ ions leaves the crystal, and afterwards are cooled down
to a quite low temperature, half of the In ions reveal a
completely ionized state, In$^{3+}$, without any valence electron,
and the other half exhibits an In$^{1+}$ state, with two valence
electrons \cite{shcheulin1,kazanskii}. Thus the non-local
neutrality of the crystals is satisfied, and almost full
self-compensation of the donor impurities is realized. Since, due
to the Coulomb interaction, a localization of two electrons at one
orbital is energetically not profitable, a compensating local
lattice distortion appears around the In$^{1+}$ ions, the ions
being moved along a 4th order axis into a neighboring cell of
eight F$^{1-}$ ions not occupied by a Cd ion \cite{park}.
Consequently, a transition of the In ion from In$^{3+}$ state into
the In$^{1+}$ state requires a local lattice distortion. In other
words, there is a significant potential barrier between the
In$^{1+}$ and In$^{3+}$ states. Thus, In$^{1+}$ impurities in
CdF$_{2}$ form a deep level, which is similar to the
\textit{DX}-centers in typical semiconductors \cite{chadi,thio}. A
fraction of electrons is captured by In$^{3+}$ ions, forming a
hydrogen state (In$^{3+}$ + \textit{e$_{hydr}$}), a shallow donor
level, being the basic reason that the CdF$_{2}$:In reveals
semiconducting properties. The relative concentrations of the
shallow In$^{3+}$ + \textit{e$_{hydr}$} and the deep In$^{1+}$
centers depend on the temperature. Fig.\,\ref{fig1} shows an
energy level diagram for the deep, $E_{deep}$, and the shallow,
$E_{sh}$, states of the bistable In centers in CdF$_{2}$ as
function of the configuration coordinate $Q$ (position of the In
impurity relative to the surrounding ions). For a transition from
one state to another the electron has to overcome an additional
capture barrier, $E_{cap}$, and consequently the upper state is
metastable. The energy of the states measured from the bottom of
the conducting band are $E_{deep} = 0.25$ eV \cite{park} and
$E_{sh} \cong 0.1$ eV \cite{langer1}. The electrons of the
\textit{DX}-centers can be transferred to the shallow state either
by light irradiation or by temperature. The changing of the
\textit{DX}-center state leads to a change of polarizability and,
consequently, to a local change of the refractive index. This fact
allows to use the metastable shallow states to write reversible
phase holograms. The efficient writing of such holograms in the
semiconducting CdF$_{2}$ doped with In or Ga has been demonstrated
in \cite{ryskin1} and \cite{suchocki}. The interest in these
materials is caused mainly by their holographic application.

Traditionally, investigations of the semiconducting transport
properties of doped CdF$_{2}$ were carried out using ohmic
contacts. The technique to produce such contacts is quite
complicated and often is a kind of art. In this paper we report on
the radio-frequency investigations of CdF$_{2}$: In which were
carried out without a use of ohmic contacts. We have determined
characteristic features of CdF$_{2}$: In, as the temperature
dependencies of the \textit{dc} conductivity and of the ion
concentration on the donor level, the activation energy of
impurities, $E_{a}$, the total concentration of the active In
ions, $N$, the barrier height, $E_{cap}$, and the values of the
kinetic coefficients which determine the speed of the thermally
induced transfers between the deep and the shallow states of the
In ions in the CdF$_{2}$ matrix. All these parameters have been
defined from low frequency ($10 - 10^{6}$ Hz) measurements of the
complex impedance of thin plane-parallel CdF$_{2}$: In samples
with metallic electrodes which were either sputtered on their
surfaces or just brought into contact with the surfaces. This
method is the basis of our previous study \cite{shcheulin2}.

\section{Samples and Experiment}

In most of the experiments we used a homogeneous transparent
plane-parallel sample of CdF$_{2}$ with an InF$_{3}$ concentration
in the raw material of 0.02 mole \%. This concentration of the In
impurities gives the sample a red-brownish color. The absorption
coefficient $\alpha$ for light with a wavelength of $\lambda$ =
488 nm has been measured to be of the order of 50 cm$^{-1}$ at
room temperature. Control experiments have been also performed for
a pure CdF$_{2}$ sample without In doping. Both surfaces of the
sample have been covered by gold electrodes made by plasma
sputtering. The area of each electrode was 20 mm$^{2}$ and the
electrode thickness was about 10 nm. The light transmittance
through this electrode for $\lambda$ = 488 nm has been estimated
as high as 50 \%. For the experiments, where we used electrodes
isolated from the sample by mica, the same electrodes have been
sputtered on two pieces of mica, 35 $\mu$ thick each. For the
experiments with Teflon isolating linings we used polished brass
electrodes of 18.5 mm$^{2}$. Finally, additional measurements were
also performed for a sample with contacts formed by silver paint.
In all cases lead wires were glued to the electrodes by a
conducting glue.

The sample was placed on a copper cold finger of a helium flow
cryostat \textit{Helix CTI Cryogenic model 22}. The temperature of
the finger was monitored and controlled by a \textit{Lake Shore
330} temperature controller with an accuracy of 0.01 K.

In order to illuminate the samples a plane Plexiglas window was
made in the vacuum shield of the cryostat. An argon laser beam
with a power up to 14 mW at $\lambda$ = 488 nm was widened by a
lens to illuminate the complete sample surface. The laser
intensity could be continuously regulated by changing the gas
discharge current in the laser tube, and by an external variable
attenuator.

The lead wires from the electrodes were soldered to terminals of
the cryostat. The terminals were connected with an LCR meter
\textit{HP4284A}. The analyzer \textit{HP4284A} covers a frequency
range from 20 to 10$^{6}$ Hz, an amplitude \textit{ac} voltage
$V_{s}$ from 0 to 20 V and a bias voltage from - 40 to 40 V.
Usually $V_{s}$ = 0.1 V and zero bias have been used. Results of
the complex impedance of a sample with electrodes, $Z$, were
obtained as an equivalent capacitance $C_{p}$ and an equivalent
conductance $G_{p}$ (i.e. the complex conductance is 1/$Z$ =
$G_{p} + i \omega C_{p}$, where $\omega$ is the angular
frequency). Additional experiments have been carried out for
frequencies extending to 320 MHz. In these experiments an
\textit{HP4191A} impedance analyzer with working range of 1 MHz -
1 GHz has been utilized. For the measurements the sample was
placed at the end of a coaxial line, and a calibration procedure
with three standard loads was required to eliminate the
contributions of the line itself \cite{boehmer}.

\section{Results and discussion}

\subsection{Schottky barriers and a
non-homogenous layered Maxwell-Wagner capacitor model}

Typical results of our measurements of the frequency dependencies
of $C_{p}$, and $G_{p}/\nu$ for the sample with gold electrodes
are plotted in Fig. 2 ($\nu = \omega/2\pi$). These measurements
were performed with the sample cooled in darkness. They clearly
show the signature of a relaxation process for frequencies 20 Hz
to 1 MHz, the relaxation frequency $\nu_{p}$ strongly depending on
the temperature. When the temperature changes from 110 K to 200 K,
$\nu_{p}$ rises by a factor of 10000.

\begin{figure}[]
\centering
\includegraphics[width=\columnwidth,clip]{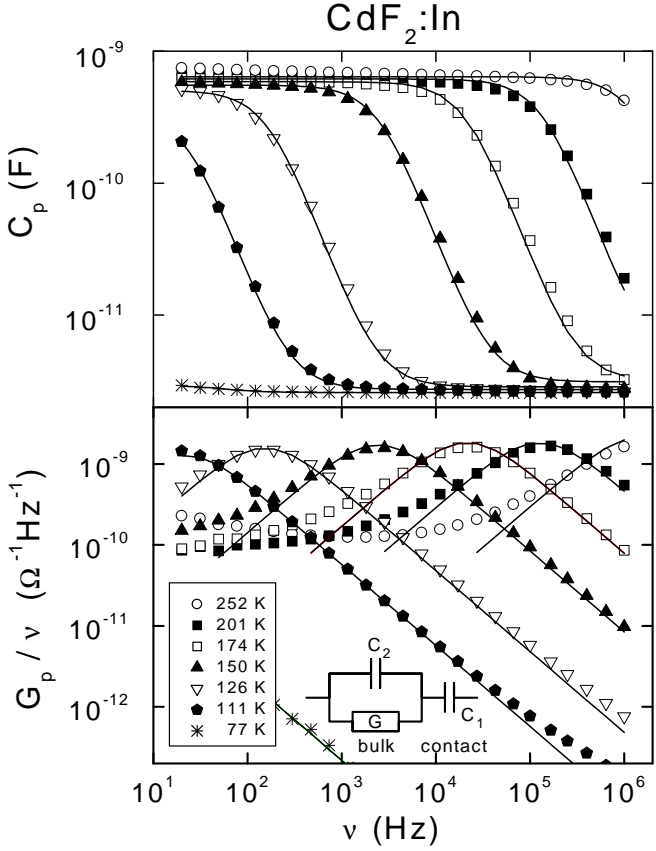}
%\vspace{0.2cm}
\caption{Frequency dependencies of $C_{p}$ (upper panel) and
$G_{p}/\nu$ (bottom panel) for a sample with gold-sputtered
electrodes at various temperatures. The points are experimental
data, the solid lines are least square fits by formulas \ref{gp}
and \ref{cp}, deduced from the equivalent circuit indicated in the
lower panel.} \label{fig2}
\end{figure}

Since in a cubic electronic semiconductor like CdF$_{2}$: In it is
very difficult to imagine a low frequency microscopic mechanism
for this relaxation, we propose, that a macroscopic Maxwell-Wagner
relaxation process accounts for this behavior. (This statement has
been strongly supported by the experiments described below, where
isolating layers between sample and electrodes have been used.)
The Maxwell-Wagner relaxation is a quite common feature for a
non-homogenous layered capacitor \cite{hippel}. The layered
structure results from the formation of Schottky barriers in the
regions of semiconducting CdF$_{2}$: In close to the metal
electrodes \cite{kittel}. If the electron work function
$\varphi_{m}$ in a metal is higher than in an electronic
semiconductor, $\varphi_{s}$, the electron concentration in the
contact region of the semiconductor is suppressed, and a depletion
layer appears. Its thickness is equal to
\begin{equation}
d_{c}=\sqrt{2\varepsilon \varepsilon _{0}V_{c}/en_{d}},
\label{dc}
\end{equation}
where $e$ is the electron charge, $V_{c} = (\varphi_{m} -
\varphi_{s})/e$ is the contact potential difference, $n_{d}$ is
the donor concentration (total ionization of donors is supposed),
and $\varepsilon$ and $\varepsilon_{0}$ are the dielectric
constants of the semiconductor and of the vacuum.

\begin{table*}
\caption{\label{table1}Experimental values of $\varepsilon$ and
$\sigma$ of the bulk material obtained from the radio-frequency
measurements with three different kinds of linings at the
electrodes: 1) Teflon linings, 2) mica linings, and 3) without any
linings (marked as "Schottky barrier"). The values are presented
for three temperatures. The experimental accuracy for
$\varepsilon$ is: 30 \% for the Teflon linings, 20 \% for the mica
linings, and 10 \% for the measurements without linings. The
experimental accuracy for $\sigma$ is: 30 \% for the Teflon
linings, 10 \% for the mica linings, and 4 \% for the measurements
without linings.}

\begin{ruledtabular}
\begin{tabular}{c|ccc|ccc|ccc}
 &\multicolumn{3}{c|}{Teflon}&\multicolumn{3}{c|}{Mica}&\multicolumn{3}{c}{Schottky barrier}\\
 T(K)&109&124&148&109&124&148&110.5&125.3&149.4\\ \hline
 $\varepsilon$&8.7&9.4&10.9&10.2&10.3&11.0&7.3&7.7&8.3 \\
 $\sigma$($\Omega^{-1}$cm$^{-1}$)&$1\cdot10^{-8}$&$1\cdot10^{-7}$&$1\cdot10^{-6}$&$1.4\cdot10^{-8}$&
 $1.3\cdot10^{-7}$&$2.2\cdot10^{-6}$&$1.3\cdot10^{-8}$&$1.2\cdot10^{-7}$&$2.2\cdot10^{-6}$\\
\end{tabular}
\end{ruledtabular}
\end{table*}

Thus, the sample with two electrodes may be considered as a
structure of three condensers in series: two of them are formed by
the depletion layers with small conductivity, and the third (the
middle) one represents the bulk material with the real sample
conductivity. If the material of both electrodes are the same and
both sample surfaces had the same treatment, then two
depletion-layer capacitors with capacity $C^{\prime}$ may be
presented in the equivalent circuit scheme as one capacitor with
capacitance $C_{1} = C^{\prime}/2$. The rest of the sample may be
considered as a capacity $C_{2}$ with a parallel active
conductivity $G$. Overall we arrive at the equivalent circuit
indicated in Fig. 2. Neglecting the conductivity of the depletion
layers, the impedance of this scheme is:
\begin{equation}
Z=\frac{1}{i\omega C_{1}}+\frac{1}{G+i\omega C_{2}},  \label{zzz}
\end{equation}
and the complex conductance is $1/Z = G_{p} + i \omega C_{p}$,
where
\begin{equation}
G_{p}=\frac{C_{1}^{2}\omega ^{2}G}{\omega
^{2}(C_{1}+C_{2})^{2}+G^{2}}, \label{gp}
\end{equation}
\begin{equation}
C_{p}=\frac{C_{1}\left[ G^{2}+\omega ^{2}(C_{1}C_{2}+C_{2}^{2})\right] }{%
\omega ^{2}(C_{1}+C_{2})^{2}+G^{2}}.  \label{cp}
\end{equation}
An analysis of (3) and (4) shows that at $\omega\rightarrow0$
\begin{equation}
C_{p}\rightarrow C_{p0}=C_{1},  \label{cp0}
\end{equation}
\begin{equation}
G_{p}\rightarrow G_{p0}=0;  \label{gp0}
\end{equation}
and at $\omega\rightarrow\infty$
\begin{equation}
C_{p}\rightarrow C_{p\infty }=C_{1}C_{2}/(C_{1}+C_{2}),
\label{cpinf}
\end{equation}
\begin{equation}
G_{p}\rightarrow G_{p\infty }=GC_{1}^{2}/(C_{1}+C_{2})^{2}.
\label{gpinf}
\end{equation}
If $C_{1}\gg C_{2}$ (that usually holds for the Schottky
barriers), then:
\begin{equation}
C_{p\infty }=C_{2}, \label{cpinf1}
\end{equation}
\begin{equation}
G_{p\infty }=G. \label{gpinf1}
\end{equation}
Thus, the high frequency limit represents the bulk material
parameters, while the low frequency one depends strongly on the
contact phenomena. In addition, $C_{p}(\omega _{p})=(C_{p\infty
}+C_{p0})/2$, where the relaxation frequency is:
\begin{equation}
\omega _{p}=2\pi \nu _{p}=G/(C_{1}+C_{2}).  \label{omega}
\end{equation}
The frequency dependencies of $C_{p}$, and $G_{p}/\nu$ calculated
from formulas \ref{gp} and \ref{cp} are shown in Fig. \ref{fig2}
as solid lines. The parameters $C_{1}$, $C_{2}$ and $G$ have been
fitted with the least square method. As can be seen from Fig.
\ref{fig2}, all the characteristic features of the experimental
curves are well described by the formulas of the Maxwell-Wagner
model \cite{hippel}. For the higher temperatures, at low
frequencies deviations of experimental data and fits show up in
$G_{p}/\nu$. They can be ascribed to a small, but non-zero
conductivity of the depletion layers, which for very low
frequencies would lead to a $1/\nu$ divergence. We note, that
$C_{p0}$ and $C_{p\infty }$ are almost temperature independent.
Consequently, the strong temperature dependence of the relaxation
frequency $\nu_{p} = G/2\pi(C_{1}+C_{2})$ is due to the
temperature dependence of $G$. Using the temperature dependencies
of the fitting parameters $C_{2}(T)$ and $G(T)$, we have
calculated the real part of the dielectric permittivity
$\varepsilon^{\prime}(T) = C_{2}(T)d/(\varepsilon_{0}S)$ and the
conductivity $\sigma(T) = G(T)d/S$ of the bulk material ($S$ is
the electrode area, and $d$ is the sample thickness).

Since the frequency dependencies of $C_{p}(\omega)$ and
$G_{p}(\omega)$ (Eqns. \ref{gp} and \ref{cp}) have the same shape
as the frequency dependencies of $\varepsilon^{\prime}$ and
$\sigma = \varepsilon^{\prime\prime}\varepsilon_{0}\omega$ for
homogeneous dielectrics with Debye relaxation (here
$\varepsilon^{\prime}$ and $\varepsilon^{\prime\prime}$ are real
and imaginary parts of the dielectric permittivity), we have
conducted two control experiments for the CdF$_{2}$: In sample
with electrodes isolated from the sample by 55 $\mu$ thick Teflon
($\varepsilon_{Teflon}$ = 2) and 35 $\mu$ thick mica
($\varepsilon_{mica}$ = 8) layers. In these experiments no
Schottky barriers are formed, and one can directly calculate the
electrode capacitances $C^{\prime} = 2C_{1} \approx 2C_{p0}$ of
the Maxwell-Wagner layered system. We have found, that the value
of $C_{1}$ for Teflon is two orders of magnitude smaller and for
mica it is one order of magnitude smaller than $C_{p0}$ in the
experiments with sputtered gold electrodes (2.5 pF and 20 pF,
respectively), while $C_{2}$ remains at 3 - 4 pF. Accordingly, for
a fixed temperature the values of $C_{p0}$ decrease and the
characteristic relaxation frequencies increase by 2 (1) orders of
magnitude for Teflon (mica) linings. Such a behavior is expected
in the case of a Maxwell-Wagner relaxation (see Eqn. \ref{omega}),
since the bulk conductance $G$ does not depend on the type of the
contacts. At the same time, in the case of the Debye relaxation,
the characteristic frequency of which does not depend on the
electrode capacitances, one would obtain other, more complex,
frequency dependencies of $C_{p}$ and $G_{p}$ due to a
superposition of the Maxwell-Wagner relaxation and the Debye
relaxation.

The evaluated values of $\varepsilon^{\prime}$ and $\sigma$ of the
bulk material are presented in Table \ref{table1} for three
temperatures together with the values determined from the
measurements with Schottky barriers. Table 1 documents, that the
values of the dielectric constant and of the conductivity
determined through the Maxwell-Wagner equivalent circuit analysis
coincide with each other within experimental accuracy for all
temperatures. This again proves the validity of the Maxwell-Wagner
relaxation as a model for the treatment of our results.

According to the weak temperature dependence of $C_{p\infty}$, the
capacity $C_{2}$ does not depend significantly on temperature: at
temperatures between 70 K and 300 K $C_{2} \approx 3$ pF, and
consequently $\varepsilon \cong 8$ (at lowest T = 25 K,
$\varepsilon \cong 7$). The same values of $\varepsilon$ have been
determined for an undoped sample. Contactless measurements made in
the submillimeter frequency range give the same values for doped
and pure CdF$_{2}$ as well \cite{pronin}.

\subsection{Temperature and frequency dependencies of the conductivity}

\begin{figure}[]
\centering
\includegraphics[width=6cm,clip]{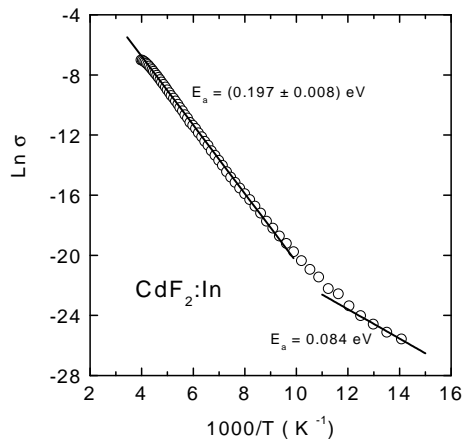}
%\vspace{0.2cm}
\caption{Arrhenius plot of the temperature dependence of the
conductivity obtained by the fitting of the experimental curves
$C_{p}(\nu)$ and $G_{p}(\nu)$ for different temperatures with the
Maxwell-Wagner model, Eqns. \ref{gp} and \ref{cp}.} \label{fig3}
\end{figure}

The temperature dependence of the conductivity obtained by fitting
the experimental results $C_{p}(\omega)$, $G_{p}(\omega)$ with the
Maxwell-Wagner model is shown in Fig. \ref{fig3}. For temperatures
100 K $\div$ 250 K the temperature dependence of the conductivity
can be well described by a thermally activated behavior:
\begin{equation}
\sigma(T) \propto exp(-E_{a}/k_{B}T),  \label{sigma}
\end{equation}
with an activation energy $E_{a} = 0.197 \pm 0.008$ eV. This
value, being actually "an weighted-mean activation energy" of the
deep, $E_{deep}$, and the shallow, $E_{sh}$, states of the In
ions, takes into account both the electron transfers from the deep
and the shallow levels into the conduction band, and the process
of transferring electrons between the deep and the shallow
metastable donor levels. As can be seen from Fig. \ref{fig3}, at
low temperatures the slope of the $ln\sigma(1/T)$ curve decreases,
corresponding to the decrease in the electron replenishment of
shallow levels from the deep ones. The value of $E_{a}$,
consequently, should approach the activation energy of the shallow
state at quite low temperatures. This behavior has been indeed
observed in Ref. \cite{kunze} at temperatures 40 K to 80 K.

Since the electron mobility $\mu$ in CdF$_{2}$ depends only weakly
on temperature (at the temperatures $70 \div 300$ K, $\mu \approx
15$ cm$^{2}$V$^{-1}$s$^{-1}$ \cite{khosla}), we can calculate the
electron concentration $n_{e}$ in the conduction band. For
example, for T = 150 K one finds $n_{e} = \sigma/e\mu =
1\cdot10^{12}$ cm$^{-3}$. For this value of the electron
concentration the total effective concentration of the In ions
with the activation energy of $E_{a}$ = 0.197 eV, should be $N =
n_{e} \cdot exp(E_{a}/k_{B}T) = 4\cdot10^{18}$ cm$^{-3}$, that
coincides with $N = (3.5 \pm 0.7) \cdot 10^{18}$ cm$^{-3}$
obtained from our absorption coefficient data $\alpha \cong 50$
cm$^{-1}$ at room temperature for $\lambda$ = 488 nm.

\begin{figure}[]
\centering
\includegraphics[width=\columnwidth,clip]{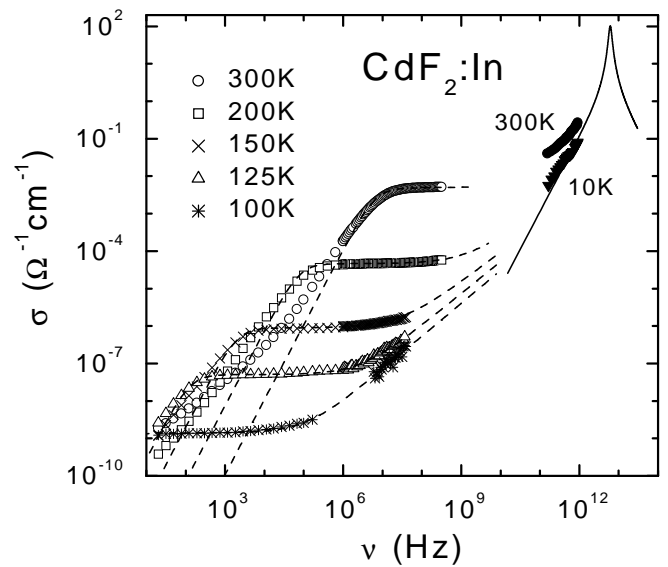}
%\vspace{0.2cm}
\caption{Panorama spectra of the conductivity of CdF$_{2}$:In with
silver-paint electrodes for various temperatures (note the log-log
scale). Open symbols - audio and radio-frequency measurements,
solid symbols - submillimeter wave measurements \cite{pronin},
solid line - infrared measurements of undoped CdF$_{2}$
\cite{axe}. Dashed lines are calculated with the equivalent
circuit shown in Fig. \ref{fig2} with an additional UDR element
for the bulk response \cite{lunkenheimer, sichelschmidt}. }
\label{fig4}
\end{figure}

Fig. \ref{fig4} compares our results on the low-frequency
conductivity in CdF$_{2}$:In with the measurements in the
submillimeter range carried out on a different In-doped sample at
10 K and 300 K \cite{pronin} and with room-temperature infrared
measurements in undoped CdF$_{2}$ \cite{axe}. The radio-frequency
experimental data shown in Fig. \ref{fig4}, extended up to 1 GHz,
were obtained on the same sample as the results of Fig.
\ref{fig2}, but with contacts made from silver paint. The
dielectric-loss results up to 1 MHz (not shown) are similar to
those shown in Fig. \ref{fig2}. In the conductivity representation
of Fig. \ref{fig4}, the loss peaks are transformed into a steep
increase of $\sigma(\omega)$, followed by the approach of a nearly
frequency-independent plateau value. The initial increase of
$\sigma(\omega)$ can be ascribed to the successive shorting of the
high resistance of the depletion layer by its capacitance. One
should be aware that in the contact-dominated region
$\sigma(\omega)$ shown in Fig. \ref{fig4}, which was calculated
from the conductance $G(\omega)$ using the geometry of the sample,
does not reflect the true conductivity of the sample material.
Only when the plateau region is reached, the intrinsic bulk
response is measured. At the lower temperatures, following this
plateau, $\sigma(\omega)$ starts to increase again with increasing
frequency. Such a behavior is often observed in amorphous and
doped semiconductors and usually ascribed to hopping conductivity
of localized charge carriers \cite{elliott}. It can be
parameterized by the so-called "Universal Dielectric Response"
(UDR): $\sigma  = \sigma_{dc} + \sigma_{0}\omega^{s}$, $s < 1$
\cite{jonscher}. The dashed lines in Fig. \ref{fig4} have been
calculated using the equivalent circuit of Fig. \ref{fig2}, with
an additional UDR element (including its contribution to
$\sigma^{\prime\prime}$ via the Kramers-Kronig relation
\cite{jonscher}) connected in parallel to $G$ and $C_{2}$
\cite{lunkenheimer, sichelschmidt}. In this way the general
behavior of the experimental spectra can be satisfactorily
reproduced. Values of $s$ between 0.8 and 0.88 were obtained,
which lies in a reasonable range for hopping conduction
\cite{elliott}.  The deviations showing up at low frequencies,
again indicate that there is a non-zero conductivity of the
depletion layers. For 300 K, the low-frequency curve matches well
with the submillimeter data and possibly the increase of
$\sigma(\omega)$ observed in the submillimeter region can also be
taken into account by the UDR. Up to now UDR behavior only rarely
has been observed up to such high frequencies (e.g.,
\cite{sichelschmidt}). However, as mentioned above, the
measurements in the submillimeter range were performed on a
different sample, which may have a slightly different In content
and therefore Fig. \ref{fig4} can only provide a qualitative
comparison of the low- and high-frequency response. It may be
noted that in CdF$_{2}$:In there is no indication for the
characteristic frequency dependence connected with Drude behavior:
when the frequency becomes comparable to the scattering rate of
the charge carriers, a characteristic frequency-dependent decrease
of the conductivity should show up. The absence of this feature
possibly is due to the fact that the high-frequency conductivity
seems to be governed by hopping of localized charge carriers,
while Drude-like transport of "free" electrons, excited into the
conduction band, dominates at low frequencies and \textit{dc}
only.

The submillimeter-range data for 10 K, where the low-frequency
conductivity in the doped sample can be expected to be extremely
low, agree well with the infrared data for CdF$_{2}$ without
impurities (solid line). The latter, showing a peak at around
10$^{13}$ Hz represents a dipolar lattice mode.

\begin{figure}[]
\centering
\includegraphics[width=5cm,clip]{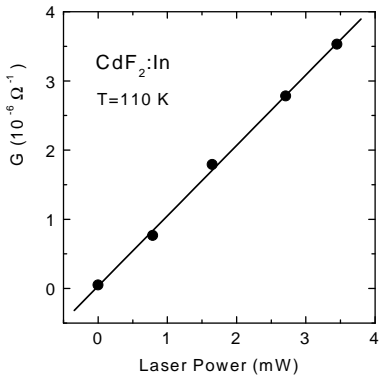}
%\vspace{0.2cm}
\caption{The CdF$_{2}$:In sample conductance $G$ versus the laser
power at $T$ = 110 K ($G$ is obtained from the fitting of
$C_{p}(\nu)$ and $G_{p}(\nu)$ curves).} \label{fig5}
\end{figure}

\subsection{Light excitation and measurements of the shallow state kinetics}

\begin{figure}[]
\centering
\includegraphics[width=\columnwidth,clip]{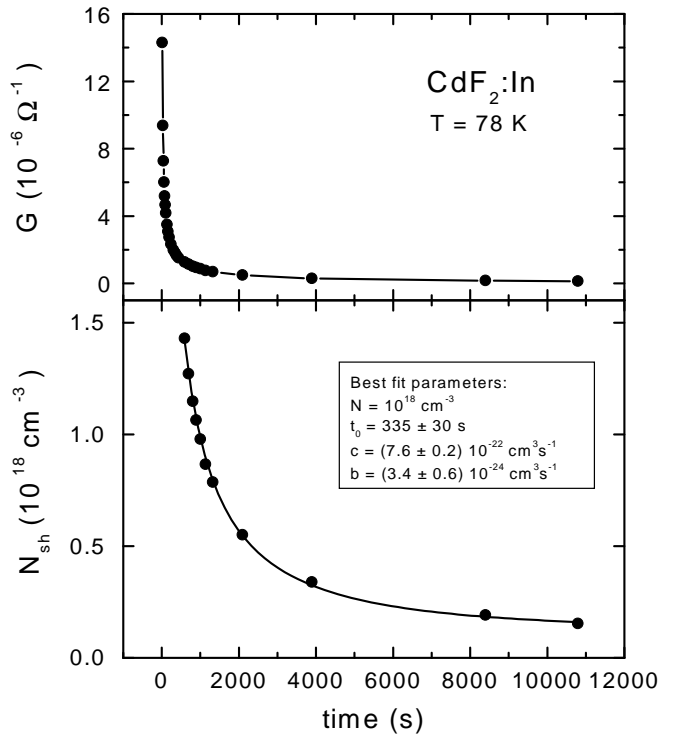}
%\vspace{0.2cm}
\caption{Time dependencies of the conductance $G$ (upper panel)
and of the shallow donor concentration $N_{sh}$ (bottom panel) of
CdF$_{2}$:In after its irradiation with 488 nm laser light at $T$
= 78 K. Upper panel: points are the experimental data, line is
guided for eye. Bottom panel: points are calculated from Eqn.
\ref{nsh1} using $n_{e}(t) = G(t)d/Se\mu$, and line is the best
fit by Eqn. \ref{nsh} (only those $G(t)$ data points have been
used here, for which the calculated $N_{sh}$ is much less than the
total concentration of active impurity $N = 4 \cdot 10^{18}
cm^{-3}$).} \label{fig6}
\end{figure}

In order to determine the kinetic coefficients which affect the
rate of the thermally induced transfers between the deep and
shallow states, we have carried out experiments with illumination
of the sample by light with a wavelength $\lambda = 488$ nm. This
wavelength hits into the wide photoionization absorption band in
CdF$_{2}$:In centered around 400 nm \cite{shcheulin1}. The sample
was cooled to the required temperatures in darkness and
illuminated afterwards. We found, that the characteristic
relaxation frequency $\nu_{p}$ is proportional to the intensity of
the light, $I_{laser}$, at all the temperatures (50 K $\div$ 150
K) at which under illumination $\nu_{p}$ hits into the
instrumental frequency window of the \textit{HP4284} analyzer.
This behavior can be ascribed to the conductivity being
proportional to the light intensity (at least for that used in our
experiments $I_{laser} < 6$ mW/cm$^{2}$), since at $C_{1} \cong
const$ and $C_{2} \cong const$ the conductance of the sample is $G
= 2\pi \nu_{p}(C_{1}+C_{2}) \propto I_{laser}$ (cf. Eqn.
\ref{omega}). As an example, Fig. \ref{fig5} shows the sample
conductance versus the laser power for $T$ = 110 K. The linear
dependencies $G(I_{laser})$ show the absence of any saturation
effects for the illumination power used in our experiments. After
switching off the illumination, the frequency $\nu_{p}$ and the
conductance $G$ relax with time. The rate of this relaxation is
not constant. During first minutes $G$ rapidly drops by 1 or 2
orders of magnitude, but then decreases rather gradually with a
characteristic time being from half an hour to several hours (Fig.
\ref{fig6}) depending on temperature. This process is connected
with the capture of a part of the non-equilibrium carriers by the
In$^{3+}$ ions into the metastable donor state In$^{3+} +
e_{hydr}$ with a subsequent slow transition into the ground state
In$^{1+}$. Heating the sample to room temperature and subsequent
recovery cooling in darkness completely restores $\nu_{p}$ and the
frequency dependencies of $C_{p}(\nu)$ and $G_{p}(\nu)$.

It is shown in Ref. \cite{shcheulin1}, that during illumination of
CdF$_{2}$:In by light with a wavelength inside of the
photoionization absorption band, the following photochemical
reaction takes place:
\begin{equation}
In^{1+} + In^{3+} + h\nu \rightarrow 2(In^{3+} + e_{hydr}).
\label{photoionization}
\end{equation}
This reaction brings the In ions into the hydrogen-like state with
an energy $E_{sh}$. The reverse process of the thermal decay of
the shallow centers with capture of the released electrons by
other shallow centers occurs in accordance with the bimolecular
reaction:
\begin{equation}
2(In^{3+} + e_{hydr}) + kT_{B} \rightarrow In^{1+} + In^{3+}.
\label{bimolecular}
\end{equation}
The thermal decay of the shallow states after the light
illumination is described by the kinetic equation
\cite{shcheulin1}:
\begin{equation}
dN_{sh}/dt = - cN_{sh}^{2} + b(N - N_{sh})^{2}/4,
\label{thermaldecay}
\end{equation}
where $N_{sh}$ is the concentration of the shallow centers, $c$
and $b$ are the kinetic coefficients of the decay and the creation
of the shallow center, correspondingly. It is assumed here, that
the concentration $N^{1+}$ of the deep centers is approximately
equal to the concentration $N^{3+}$ of the completely ionized
In$^{3+}$ centers, $N^{1+} + N^{3+} + N_{sh} = N$ and,
consequently, $N^{1+} \approx N^{3+} \approx (N - N_{sh})/2$ and
the free electron concentration is negligible. The solution of the
Eqn. \ref{thermaldecay} yields the following time dependence for
$N_{sh}$:
\begin{equation}
N_{sh}(t) =
\frac{n_{1}exp\left[N\sqrt{cb}(t+t_{0})\right]-n_{2}}{exp\left[N\sqrt{cb}(t+t_{0})\right]-1},
\label{nsh}
\end{equation}
$n_{1} = N/(1 + 2\sqrt{c/b})$, $n_{2} = N/(1 - 2\sqrt{c/b})$,
$t_{0}$ is a constant of integration.

In order to compare our results for $G(t)$ with Eqn. \ref{nsh} we
have calculated the electron concentration $n_{e}(t) = G(t)d
/(Se\mu)$, and then obtained $N_{sh}(n_{e})$, using the procedure
described below.

Since in the CdF$_{2}$:In crystals, after the additive coloration
process, the In$^{3+}$ ions play a  role of donors, and the
In$^{1+}$ ions, capturing an additional electron, play a role of
acceptors, and since the effective concentration of donors $n_{d}$
in a compensated donor semiconductor is equal to the difference
between the concentrations of donors and acceptors \cite{smith},
we obtain $n_{d}$ equal to the concentration of the shallow
hydrogen-like centers $N_{sh}$, formed via capture of electrons to
a hydrogen-like orbit by a little part of the In$^{3+}$ ions. We
assumed thermal equilibrium between the conduction electrons and
the shallow states. We also assume that all the conduction
electrons are formed through ionization of the shallow states and
that the direct electron transitions between the deep levels and
the conduction band are negligible. The conducting electron
concentration in a semiconductor with donor and acceptor
impurities with concentrations of $N_{d}$ and $N_{a}$ ,
correspondingly, is \cite{smith}:
\begin{equation}
N_{e} = \frac{N_{d} - N_{a}}{2N_{a}}exp(-E_{d}/k_{B}T),
\label{smith}
\end{equation}
where $E_{d}$ is the energy of the donor level, $N_{c} = 2(2\pi
m^{*}k_{B}T)^{3/2}/h^{3}$ is the density of states in the
conduction band and $m^{*}$ the effective electron mass. Eqn.
\ref{smith} is valid for low enough temperatures, when
\begin{equation}
E_{d}/k_{B}T \gg 1. \label{ed}
\end{equation}
In our case $N_{d} \equiv N^{3+}, N_{a} \equiv N^{1+}$ and  $N_{d}
- N_{a} = N_{sh}$, $N^{3+} \approx N^{1+} \approx N/2$, $E_{d}
\equiv E_{sh} = 0.1$ eV and the condition of Eqn. \ref{ed} is
always satisfied. Consequently, the electron concentration is
given by:
\begin{equation}
n_{e} = \frac{N_{sh}}{N}N_{c}exp(-E_{sh}/k_{B}T),  \label{ne}
\end{equation}
and
\begin{equation}
N_{sh}(t) = (n_{e}(t)N/N_{c})exp(E_{sh}/k_{B}T).  \label{nsh1}
\end{equation}
Substituting to Eqn. \ref{nsh1} the data for $N$ and $n_{e}(T)$,
defined in our experiments ($N = 4\cdot10^{18}$ cm$^{-3}$,
$n_{e}(T) = \sigma(T)/e\mu$), and the values of $m^{*} = 0.45
m_{e}$ \cite{khosla} and $E_{sh} = 0.1 eV$, we obtain
$N_{sh}/n_{e} = 1.15\cdot10^{7}$ at T = 78 K and $N_{sh}/n_{e} =
9.07\cdot10^{4}$ at T = 110 K. In accordance with assumed thermal
equilibrium between the conduction electrons and the shallow
states we use these ratios for calculating $N_{sh}(t)$ through the
experimentally found time-dependent electron density $n_{e}(t) =
G(t)d/Se\mu$. The strong identity of the time dependencies of the
infrared absorption ($\propto N_{sh}(t)$) and of the conductivity
($\propto n_{e}(t)$) after an optical excitation has been
experimentally shown in \cite{kunze}. Since Eqn.
\ref{thermaldecay} implies that $N_{sh} \ll N$ , we used only
those experimental $G(t)$ data points, for which the calculated
$Nsh$ is much less than $4\cdot10^{18}$ cm$^{-3}$.

The results for $N_{sh}(t)$ and the calculated fit curve Eqn.
\ref{nsh} are shown in Fig. \ref{fig6} (bottom part) for T = 78
with the best fit parameters ($N = 10^{18}$ cm$^{-3}$, $t_{0} =
335$ s, $c =7.6\cdot10^{-22}$ cm$^{3}$s$^{-1}$, and $b =
3.4\cdot10^{-24}$ cm$^{3}$s$^{-1}$). For the thermal decay curve
at T = 110 K the best fit parameters are $N = 4\cdot10^{18}$
cm$^{-3}$, $t_{0} = 2875$ s, $c =1.17\cdot10^{-19}$
cm$^{3}$s$^{-1}$, and $b = 3.7\cdot10^{-25}$ cm$^{3}$s$^{-1}$.
Using these values of $c$ and $b$, we compare the shallow center
decay and creation processes for $T$ = 78 K and $T$ = 110 K. At
$t$ = 1000 s and $T$ = 78 K, the first term of Eqn.
\ref{thermaldecay} is two orders of magnitude larger than the
second one, i.e. the thermal decay of the shallow states is
dominating.  At T = 110 K and $t$ = 1000 s these terms are nearly
equal, i.e. the thermal decay begins to be compensated by the
thermal activation. It is interesting, that $c$(110 K) is two
orders of magnitude larger than $c$(78 K), while $b$(110 K) is one
order of magnitude smaller than $b$(78 K).

Assuming that the temperature dependence of the kinetic
coefficient $c$ can be described by activated behavior, $c(T) = A
\cdot exp(-E_{ac}/k_{B}T)$, from $c$(78 K) and $c$(110 K) we found
$E_{ac}$ = 0.12 eV. This value coincides with $E_{ac}$ obtained in
\cite{shcheulin1} by the kinetic measurements of the shallow state
infrared absorption. The kinetic coefficient $c$ may be presented
as $c(T)= const\cdot p_{1}\cdot p_{2}$, where $p_{1}=\nu_{1} \cdot
exp(-E_{sh}/k_{B}T)$ is the rate of the electron releasing out the
first shallow center to the conduction band, and $p_{2} = \nu_{2}
\cdot exp(-E_{cap}/k_{B}T)$ is the rate of the thermally activated
electron hopping over the barrier $E_{cap}$ and simultaneous
capture of a conduction electron by the second shallow center
\cite{shcheulin1}. Here $\nu_{1}$ and $\nu_{2}$ are the attempt
frequencies: $\nu_{1}$ is the impurity vibration frequency and
$\nu_{2}$ is the "configuration phonon mode" frequency of the
impurity. The electron-impurity collision frequency $\nu_{coll}$
in our case is much higher than $\nu_{2}$: since the electron
mobility at $T \leq 150$ K is determined mainly by the ionized
impurity scattering \cite{khosla}, the first can be calculated as
$\nu_{coll} \approx e/m^{*}\mu = 2.6\cdot10^{14}$ Hz, while for
the last the upper estimate would be the Debye frequency
$\nu_{Debye} \approx 6\cdot10^{12}$ Hz. Thus, at least several
dozens of electron-impurity collisions occur during one cycle of
the impurity "configuration phonon mode" and at the moment of the
local lattice distortion near the In impurity there is always a
conduction electron ready to be captured. Thus, for the activation
energy of the kinetic coefficient $c$ one has $E_{ac} = E_{sh} +
E_{cap}$, and the capture barrier is equal to $E_{cap} = E_{ac} -
E_{sh} = 0.12$ eV - 0.1 eV = 0.02 eV.

For a higher temperature ($T$ = 150 K) the shallow state decay
curve is not fitted exactly by Eqn. \ref{nsh}. The change of the
decay kinetics type on increasing temperature may be explained by
formation of a shallow states impurity band in CdF$_{2}$:In. The
levels of the hydrogen-like states In$^{3+} + e_{hydr}$ may form
such a band due to Coulomb interaction with statistically
distributed F$^{1-}$, In$^{3+}$ and In$^{1+}$ ions \cite{kunze}.
For increasing temperature, the higher levels of this band become
populated, leading to an effective decrease of $E_{cap}$. At $T$ =
150 K the levels up to $k_{B}T$ = 0.013 eV $\approx E_{cap}$ are
populated and the barrier is eliminated.

\begin{figure}[]
\centering
\includegraphics[width=\columnwidth,clip]{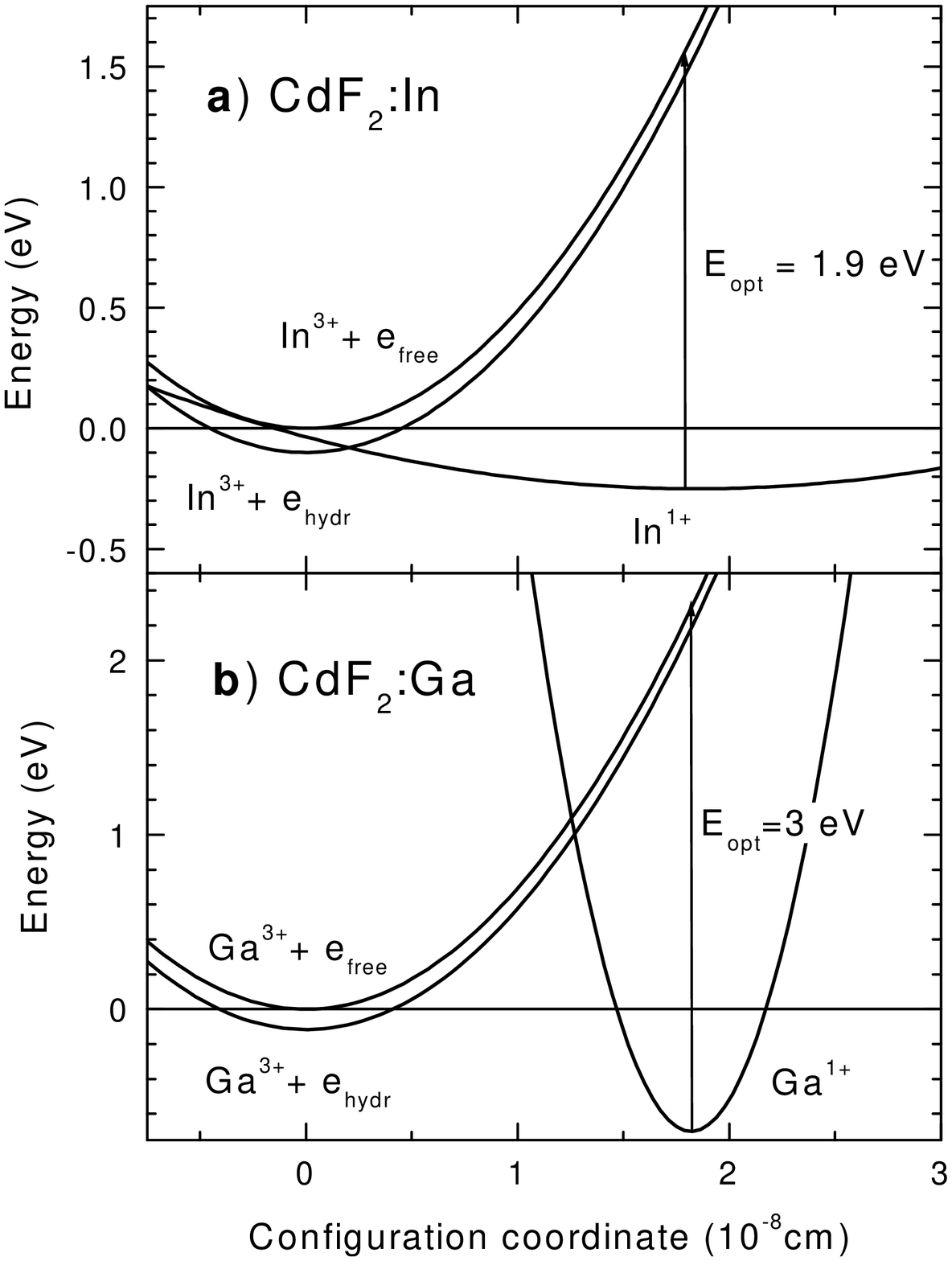}
\vspace{0.2cm} \caption{Specified energy levels diagrams for the
deep and the shallow states of the bistable In (a) and Ga (b)
center in CdF$_{2}$, as function of the configuration coordinate
$Q$ of the impurity ion nucleus.} \label{fig7}
\end{figure}

The knowledge of the key parameters $E_{cap}$ = 0.02 eV, $E_{sh}$
= 0.1 eV \cite{langer1}, $E_{deep}$ = 0.25 eV \cite{park,langer1},
$E_{opt}$ = 1.9 eV \cite{park} (the optical ionization energy of
the deep state) and $Q_{0} - Q_{1}$ = 1.84 \AA \cite{langer1} (the
difference of configuration coordinates of the deep and the
shallow states) allows to define concretely the energy diagram
(presented schematically in Fig. \ref{fig1}), assuming $Q_{1}$ = 0
and a quadratic $Q$-dependence of the levels:
\begin{equation}
\begin{array}{l}
E_{cond}(Q)=a_{1}Q^{2}, \\
E_{sh}(Q)=a_{1}Q^{2}-E_{sh}, \\
E_{deep}(Q)=a_{0}(Q-Q_{0})^{2}-E_{deep}.
\end{array}
\label{qdependence}
\end{equation}
Here the energies are measured in eV and the coordinates are in
Angstroms. Then for $a_{1}$ and $a_{0}$ one has:
\begin{equation}
\begin{array}{l}
a_{1}=\frac{E_{opt}-E_{deep}}{Q_{0}^{2}}, \\
a_{0}=\frac{E_{deep}-E_{sh}+E_{cap}}{(Q_{cap}-Q_{0})^{2}}, \\
Q_{cap}=\sqrt{\frac{E_{cap}}{a_{1}}},
\end{array}
\label{a0a1}
\end{equation}
consequently, $a_{1}$ = 0.487 eV/\AA$^{2}$, $a_{0}$ = 0.063
eV/\AA$^{2}$, $Q_{cap} = 0.20$ \AA. The corresponding diagram is
given in Fig. \ref{fig7}a, which shows that the walls of the
potential well of the shallow states are much steeper then those
of the deep state. Therefore, according to the values of $a_{1}$
and $a_{0}$, the "return force" $F = - dE/dQ$ is 8 times larger
for the shallow state. Besides, the barrier position $Q_{cap}$ is
very close to the potential well minimum of the shallow state . It
is interesting to note, that a similar calculation made for
CdF$_{2}$ doped with Ga ($E_{sh}$ = 0.116 eV \cite{langer2},
$E_{opt}$ = 3 eV, $E_{cap}$ = 1.12 eV \cite{ryskin2}, $E_{deep}$=
0.7 eV and $Q_{0} - Q_{1}$ = 1.82 \AA \cite{park}) gives the
opposite picture of the potential curves (Fig. \ref{fig7}b). Now
$a_{1}$ = 0.694 eV/\AA$^{2}$, $a_{0}$ = 5.63 eV/\AA$^{2}$,
$Q_{cap} = 1.27$ \AA, i.e. the "return force" for the shallow
state is 8 times weaker than one for the deep state, and the
barrier position $Q_{cap}$ is far away from the shallow state
potential well minimum. In addition, while the "return forces" for
the shallow states of In and Ga in CdF$_{2}$ are close to each
other, the "return force" for the deep state of Ga is 90 times
larger then one for In. Now we can calculate the impurity
"configuration phonon mode" frequencies of the deep, $\nu_{deep} =
\frac{1}{2\pi}\sqrt{2a_{0}/M}$, and of the shallow, $\nu_{sh} =
\frac{1}{2\pi}\sqrt{2a_{1}/M}$, states. Here $M$ is the reduced
mass of the impurity ion and of the surrounding ions, involved in
the "configuration phonon modes". As the first approximation we
have taken the values of the impurity ions masses for $M$ and
obtained for the In impurity $\nu_{deep} = 5.18\cdot10^{11}$ Hz,
$\nu_{sh} = 1.44\cdot10^{12}$ Hz and for the Ga impurity
$\nu_{deep} = 6.28\cdot10^{12}$ Hz, $\nu_{sh} = 2.2\cdot10^{12}$
Hz.

We should note, that the data obtained for the values of kinetic
coefficients are at some extent qualitative, because the
assumption that the direct transitions between the deep levels and
the conduction band is negligible is not strongly fulfilled in
CdF$_{2}$:In at temperatures between 78 and 150 K. Nevertheless,
the used ratios $N_{sh}/n_{e}$ calculated by the simple
approximate Eqn. \ref{nsh1} for $T$ = 78 K and $T$ = 110 K
coincided within a few tenth of a percent with the ratios obtained
with the formulae of exact statistical calculation derived in
\cite{shcheulin2}. Similar measurements and data analysis for the
CdF$_{2}$:Ga crystals would give accurate results, since in this
crystal $E_{deep}$ = 0.7 eV \cite{park}, and the assumption
mentioned above is fulfilled. Unfortunately, the CdF$_{2}$:Ga
samples reveal often some amount of the In impurities as well
\cite{koziarska}, influencing the experimental results, since
there are actually two activation energies and two capture
barriers in these samples.

\section{Conclusions}

In this paper we report on the low frequency conductivity
measurements of semiconducting CdF$_{2}$:In crystals with
metal-coating electrodes, producing Schottky barriers at the
sample surface. The results allow to determine a whole lot of the
material characteristics: the temperature dependence of the
\textit{dc} conductivity, the activation energy of the impurity
$E_{a}$, the total concentration of the active In ions $N$, the
shallow donor concentration $N_{sh}$, the height of the capture
barrier $E_{cap}$, and the values of the kinetic coefficients
determining the rate of the thermally induced transitions between
the deep and shallow states of In in the CdF$_{2}$ matrix. These
measurements do not require ohmic contacts, preparation of which
is necessary for standard \textit{dc} measurements and often is
quite complicated and not always realizable task.

\section{Acknowledgements}

The work was supported by BMBF (contract 13N6917/0 - EKM), CRDF
(project RP1-2096) and RFBR (project 99-02-16859).

\end{document}